\documentclass[aps,pra,floatfix,noshowkeys]{revtex4}
\usepackage{graphicx}
\usepackage{amsmath}
\usepackage{amsfonts}
\usepackage{amssymb}

\newcommand{\newc}{\newcommand}
\newc{\beq}    {\begin{equation}}
\newc{\eeq}    {\end{equation}}

\newc{\beqa}    {\begin{eqnarray}}
\newc{\eeqa}    {\end{eqnarray}}
\newc{\bs}    {\section}
\newc{\no}    {\\ \nonumber}

\def\apj{{ Astrophys. J.  }}
\def\apjl{{ Astrophys. J. Lett. }}

\def\jcap{{JCAP }}

\def\bx{{\bf{x}}}

\newcommand{\bea}{\begin{eqnarray}}
\newcommand{\eea}{\end{eqnarray}}

\newc{\st}    {\stackrel}

\begin{document}

\title{ Quantum Scales of Galaxies from Ultralight Dark Matter}

\author{Jae-Weon Lee}
%\email{scikid@kias.re.kr}
\affiliation{ Department of Electrical and Electronic Engineering, Jungwon University, Jungwon university,
            85 Munmu-ro, Goesan-eup, Goesan-gun, Chungcheongbuk-do,
              28024, Korea}

%\date{\today}
\begin{abstract}
We propose that the ultralight dark matter (ULDM) model, in which dark matter particles have a tiny mass of $m=O(10^{-22})eV$, has characteristic scales for physical quantities of observed galaxies such as mass, size, acceleration, mass flux, and angular momentum from quantum mechanics.  The typical angular momentum per dark matter particle is $\hbar$ and the typical physical quantities are functions of specific angular momentum  $\hbar/m$ and average background density of the particles. If we use the Compton wavelength instead for the length scale, we can obtain bounds for these physical quantities. For example, there is an upper bound for acceleration of ULDM dominated objects, $a_c={c^3 m}/{\hbar}$. We suggest
that the physical scales of galaxies
depend on the time of their formation and  that these  characteristic scales are related to some  mysteries of observed galaxies.
Future observations from the James Webb Space Telescope and NANOGrav can provide evidences for the presence and evolution of these scales.
\end{abstract}

%\pacs{ 98.62.Gq, 95.35.+d, 98.8O.Cq}
\maketitle

\section{Introduction}

Recently,  there is a growing interest  in
 the Ultralight Dark Matter (ULDM) model as a compelling alternative to Cold Dark Matter (CDM). In this model, dark matter particles possess an exceptionally small mass, typically on the order of $10^{-22}$ eV, and they exist in a Bose-Einstein condensate (BEC) state. (For a review see
\cite{2009JKPS...54.2622L,2014ASSP...38..107S,2014MPLA...2930002R,2011PhRvD..84d3531C,Marsh:2015xka,Hui:2016ltb}).
This model is known by various other names, including Fuzzy DM, BEC DM, scalar field DM, ultra-light axion, and wave-$\psi$ DM~\cite{1983PhLB..122..221B,Sin:1992bg,myhalo,1993ApJ...416L..71W, PhysRevLett.84.3037,PhysRevD.64.123528,repulsive,fuzzy,corePeebles,Nontopological,Mielke2009174,PhysRevD.62.103517,Alcubierre:2001ea,2012PhRvD..86h3535P,2009PhRvL.103k1301S,Fuchs:2004xe,Matos:2001ps,0264-9381-18-17-101,PhysRevD.63.125016,Julien,Boehmer:2007um,Eby:2015hsq}.
Because of its extremely small mass, ULDM particles exhibit an exceedingly high number density, resulting in a significant overlap of their wave functions, and hence they are described by macroscopic wave functions.

Galaxies, the smallest celestial objects dominated by DM, exhibit a distinctive behavior in the context of the ULDM model, where the quantum pressure arising from the uncertainty principle acts as a counterforce, preventing the collapse of dark matter halos. Therefore, the de Broglie wave length $\lambda_{dB}=\hbar/mv$
of ULDM determines the typical size of galaxies, where
$v$ is the typical velocity of the halo dark matter. In this sense
one can say that  quantum mechanical scale $\lambda_{dB}$ decides the size of galaxies. It was conjectured that this scale can explain the size and mass evolution of galaxies  ~\cite{Lee:2008ux,2009Natur.460..717V}.

Numerical simulations of galactic halos with ULDM ~\cite{Schive:2014dra} revealed that
a solitonic core exists in the halos, surrounded by granules. Both the core and granules have a characteristic size of about $kpc$. The characteristic length scale inherent to ULDM has been recognized for its ability to address several of the small-scale issues associated with CDM, including the core-cusp problem, the satellite galaxy problem and the missing satellite problem
~\cite{2022JCAP...12..033P,Salucci:2002nc,navarro-1996-462,deblok-2002,crisis}.
It is natural to investigate the characteristic scales of other physical properties such as angular momentum and acceleration in the ULDM model.

In this paper, we propose a typical scales for various physical quantities of galaxies in this model and suggest that these scales are related to mysteries of galaxies.
We provide an easy way to calculate these scales from parametrization of variables.
In the section II the characteristic scales are derived and briefly compared to the observed values.
In the section III we consider the time evolution of the scales. In the section IV we discuss the results.

\section{Scales of Fuzzy dark matter}

  ULDM particles can be described by
 a macroscopic wave function $\psi$ satisfying the following  Schr$\ddot{o}$dinger-Poisson equation (SPE);
\beqa
\label{spe}
i\hbar \partial_{{t}} {\psi} &=&-\frac{\hbar^2}{2m} \nabla^2 {\psi} +m{V} {\psi}, \no
\nabla^2 {V} &=&{4\pi G} \rho,
\eeqa
where the
 the DM mass density $\rho=mN|\psi|^2$, and $V$ is the gravitational potential. $N$ is the number of the DM particles in a system.
 The wave function is normalized
 as $\int |\psi|^2 d^3 \bx=1$.
Here,
%particles in the halo.
The SPE has
a soliton solution which well approximates
cores of galaxies and represents
typical structures in the ULDM models.

To derive the scales
 let us consider the following  dimensionless variables with hats by the parameterizations;
\begin{align}    \label{}
t & \equiv   t_c
\, \hat{t}   = \frac{\hbar^3}{m^3} \frac{1}{\left( G M\right)^2} \hat{t}, \\
\bx &  \equiv  \bx_c %\alpha_s
\, \hat{\bx}   = \frac{\hbar^2}{m^2} \frac{1}{G M} \, \hat{\bx},\\
\psi & \equiv   \psi_c \, \hat{\psi}   = \frac{m^3}{\hbar^3} \left( G M\right)^{\frac{3}{2}}  \hat{\psi} ,\\
V& \equiv  V_c \, \hat{V}   = \frac{m^2}{\hbar^2} \left({4\pi G M}\right)^{2}  \hat{V},
\label{scaling1}
\end{align}
which lead to a dimensionless version of the SPE apt for a numerical simulation;
\begin{align}    \label{SchEq}
i \, \partial_t  \hat{\psi}(\hat{\bx},\hat{t} ) \, & = -\frac{1}{2}\nabla^2 \hat{\psi}(\hat{\bx},\hat{t} )  + \hat{V}(\hat{\bx},\hat{t} ) \hat{\psi}(\hat{\bx},\hat{t} ),  %\nonumber
\\
 \nabla^2 \hat{V}(\hat{\bx},\hat{t} )&=\ 4 \pi \, |\hat{\psi}|^2(\hat{\bx},\hat{t} ). \label{codeEq}
%\eag
\end{align}
%Note that the normalization of the wave %function in the code space becomes $M_{\rm %tot}=  M \int d^3 \bx\,  |\psi|^2 $.
The scale parameters, such as $\bx_c$, represent the characteristic scales of a system, because numerical solutions of these dimensionless equations often have values of order unity.

For example,
  $t_c$ can  represent the typical relaxation time scale of ULDM soliton oscillations
  such as quasi-normal modes.
   Excited modes of ULDM halos decay to the ground state  over the characteristic time $t_c$
   by gravitational cooling which is a mechanism for relaxation by ejecting ULDM waves~\cite{gravitationalcooling}.
  The time scale can be written as
 \beq
  t_c = \frac{\hbar^3}{m^3} \frac{1}{\left( G M\right)^2} = 3.7\times 10^7 years \left(\frac{10^{-22}eV}{m}\right)^3\left(\frac{10^8 M_\odot}{M}\right)^2,
\eeq
where $M$ can be the mass of the core of a galaxy.

The origin of these scales can be roughly understood in the following way. The self-gravitational force of a ULDM soliton (i.e., boson star) is balanced by the repulsive force from the quantum pressure due to the uncertainty principle. By equating the de Broglie wavelength of the ULDM particles to the gravitational radius, we can obtain the equation
\beq
\lambda_{dB}=\hbar/(mv)=\hbar/(m\sqrt{GM/\lambda_{dB}}),
\eeq
which gives
\beq
\lambda_{dB}=\bx_c =\left(\frac{\hbar}{m}\right)^2 \frac{1}{ G M}= 854.8~pc\left(\frac{10^{-22}eV}{m}\right)^2 \frac{10^8 M_\odot}{  M}.
\label{dB}
\eeq
Therefore, $\bx_c$ is the gravitational Bohr radius
and the de Broigle wavelength of individual ULDM particles making the soliton.
(The usual scaling  $\hat{r}=\frac{mc}{\hbar} r$
corresponds to the case when $v=c$, that is, when we use the Compton wavelength instead of the de Broigle wavelength.)
The scales of ULDM systems are determined by two mass parameters $m$ and $M$.

Now, one may wonder how to determine the typical mass $M$ from cosmology.
If we know the average dark matter density $\bar{\rho}$ of the universe
one can argue that $M$ is of order of $\bar{\rho} \bx_c^3$.
Thus, from this relation
\beq
M\simeq \bar{\rho} \left( \frac{\hbar}{m} \right)^6
\left( \frac{1}{G M} \right)^3
\eeq
and one can obtain the typical mass scale
\beq
M \simeq G^{-\frac{3}{4}}\left(\frac{\hbar}{   m}\right)^{\frac{3}{2}} \bar{\rho}^\frac{1}{4},
\label{M2}
\eeq
which decrease as the time flows in the universe.
This mass scale  turns out to be the order of the quantum Jeans mass described in the next section.
Using this $M$, the length scale can be rewritten as a function of $m$ and $\bar{\rho}$ in turn;
\beq
\bx_c \simeq \sqrt{\frac{\hbar}{m}}(G \bar{\rho})^{-1/4},
\label{dB2}
\eeq
which implies that galaxies extend
in size as the universe expands ~\cite{Lee:2008ux,2009Natur.460..717V}.
Using Eq. (\ref{M2}) the time scale turns out to be
\beq
t_c\simeq \frac{1}{\sqrt{G \bar{\rho }}},
\eeq
which is just the order of the Hubble time at the formation of soliton with mass $M$. Thus, intrinsically this time scale has an  cosmological origin.

From the typical scales one can easily obtain other physical scales.
For example, the typical acceleration scale is given by
\beq
a_c=x_c/t_c^2 =G^3 m^4 M^3/\hbar^
4 =1.9\times 10^{-11}~meter/s^2\left(\frac{m}{10^{-22}eV}\right)^4\left(\frac{M}{10^8 M_\odot}\right)^3
\simeq  \sqrt{\frac{\hbar}{m}} (G \bar{\rho})^{3/4}.
\label{ac}
\eeq

Interestingly, this scale is similar to the Modified Newtonian dynamics (MOND) scale
$a_0=1.2\times 10^{-10}~meter/s^2$.
In Ref. \cite{Lee:2019ums} it is proposed that
MOND is an effective phenomenon of ULDM and
$a_c$ is related to the observed radial acceleration relation and the baryonic Tully-Fisher relation.
Eq. (\ref{ac}) implies that the effective MOND acceleration
scale decreases over time, which can be verified by a redshift dependency of it in the future observations.

%One can easily obtain other physical quantities as well.
The typical velocity is
\beq
v_c\equiv \bx_c/t_c=G M m/\hbar
=22.4~km/s \left(\frac{M}{10^8 M_\odot}\right) \left(\frac{m}{10^{-22}eV}\right)\simeq  \sqrt{\frac{\hbar}{m}} (G \bar{\rho} )^{1/4},
\eeq
which is similar to the typical velocity dispersion in a dwarf galaxy.
$v_c$ leads to a typical angular momentum in turn
\beq
L_c=M \bx_c v_c ={\hbar}\frac{ M}{m}
=1.1\times 10^{96}~\hbar ~\left(\frac{M}{10^8 M_\odot}\right) \left(\frac{10^{-22}eV}{m}\right)
\simeq \frac{\left(\frac{\hbar}{m}\right)^{5/2}\bar{\rho} ^{1/4}}{G^{3/4}}.
\eeq
This implies that  angular momentum of galactic DM halos usually decreases over time.
Interestingly (or obviously), the typical angular momentum per dark matter particle $L_c/N$ is just $\hbar$, because the typical number of the particles $N$ is $M/m$.
This result implies that each ULDM particle can have a quantized angular momentum.
Note that $v_c,L_c$, and $M$ scale
similarly as $\bar{\rho}^{1/4}$, while
$a_c$ decreases much faster.

On the other hand, the wave function scales as
\beq
\psi_c=\frac{m^3}{\hbar^3} \left( G M\right)^{\frac{3}{2}} =4~\times 10^{-5} ~pc^{-3/2}\left(\frac{m}{10^{-22}eV}\right)^3\left(\frac{M}{10^8 M_\odot}\right)^{3/2} \simeq {\left(\frac{\hbar}{m}\right)^{-3/4}} {(G \bar{\rho} )^{3/8}},
\eeq
which is not proportional to
$\bar{\rho}^{1/2}$ as naively expected.
Then, the typical density  $\rho_c\equiv M/\bx_c^3$ is
\beq
\rho_c=\frac{G^3 m^6 M^4}{{\hbar}^6}=0.16~M_\odot/pc^3~\left(\frac{m}{10^{-22}eV}\right)^6\left(\frac{M}{10^8 M_\odot}\right)^{4}.
\eeq

Similarly, one can define the probability current
\beq
J_c=\frac{\hbar}{m} Im(\psi_c\nabla \psi_c^*)\simeq
\frac{\hbar}{m} \frac{\psi_c^2}{\bx_c} =\frac{G^4 m^7 M^4}{{\hbar}^7}\simeq \frac{G m\bar{\rho}}{\hbar},
\eeq
which represents the   flux of ULDM particles in a galaxy.
Then, the mass flux is
\beq
MJ_c=3.66\times 10^{-6} M_\odot /pc^{2}/year \left(\frac{m}{10^{-22}eV}\right)^7\left(\frac{M}{10^8 M_\odot}\right)^{5},
\eeq
which is a DM mass flow per $pc^2$
per year in a region of a galaxy.

The scale for the gravitational potential,
\beq
V_c \  = \frac{m^2}{\hbar^2} \left({4\pi G M}\right)^{2}=8.8\times 10^{-7}c^2~\left(\frac{m}{10^{-22}eV}\right)^2\left(\frac{M}{10^8 M_\odot}\right)^{2} ,
\eeq
becomes about $c^2$
when the system becomes relativistic. For example, the soliton (a core of a galaxy) becomes relativistic when $M\simeq 10^{11} M_\odot$ for $m=10^{-22}eV$.

Another interesting observation is that
all the typical values above, except for $t_c$, are functions
of a specific angular momentum  $\hbar/m$ and $\bar{\rho}$. The dependency on  $\hbar/m$ means that
the heavier the ULDM particles are, the less quantum the halos are, as expected.
The  limit $m\rightarrow \infty$ corresponds to the CDM case ($\bx_c \rightarrow 0$).
The specific angular momentum is given by
\beq
\frac{\hbar}{m}=0.019\times \left( \frac{10^{-22}eV}{m} \right) pc^2/year,
\eeq
and usually the typical scales in the ULDM model is a function of this quantity~\cite{Hui:2016ltb}.

\section{Time evolution of the Scales }
The time dependency of the physical properties of galaxies is an important subject in astronomy.
Except for $\bx_c$ and $t_c$ all the quantities above are decreasing functions of the time.
Since the average density of the universe decreases as the time goes,
we expect we can find
more massive, compact, fast moving galaxies
in the early universe than in the present universe ~\cite{Lee:2008jp},
which seems to be consistent with recent observations
\cite{2023NatAs...7..622C}.
If future observations reveal the time evolution of these quantities of galaxies, it will provide another support for the ULDM model.

Let us find how the quantities evolve over time.
It is well known that the SPE has an interesting scaling property
\beq
\{t,\bx,\psi,\rho,V\}\rightarrow \{\lambda^{-2}t,\lambda^{-1}\\\\\bx,\lambda^{2}\psi,\lambda^{4}\rho,\lambda^{2}V\},
\eeq
where $\lambda$ is a scaling parameter.
This leads to a scaling law for other
quantities;
\beq
\{M,E,L\}\rightarrow \{\lambda M,\lambda^{3}E,\lambda L\},
\eeq
where $M$ is the mass, $E$ is the energy, and $L$ is the angular momentum of a dark matter halos.
This scaling property help us to
determine the time evolution of the scales.
During the matter dominated era
$\bar{\rho}$ scales as $(1+z)^{3}$, thus
by setting $\lambda=(1+z)^{3/4}$ we can easily
estimate the time evolution of the galactic halos as follows,
\begin{align}
\{t,\bx,\psi,\rho,V\}&\rightarrow  \{(1+z)^{-3/2}t,(1+z)^{-3/4}\bx,(1+z)^{3/2}\psi,(1+z)^{3}\rho,(1+z)^{3/2}V\}, \\
\{M,E,L\}&\rightarrow \{(1+z)^{3/4} M,(1+z)^{9/4}E,(1+z)^{3/4} L\}.
\end{align}
A rescaling by $\lambda$ is equivalent
to looking back to the corresponding redshift $z$.
Therefore, the ULDM model suggests
that the characteristic scales of physical quantities of galaxies
depend on the time when they formed.
The evolution of these quantities can be verified by future observations by, for example, space telescopes like James Webb space telescope.
Using the above relation
the observed size evolution of very massive compact galaxies in the early
universe can be explained, if dark matter is ULDM \cite{Lee:2008jp}.
Fig. 1 shows  the time evolution of the various typical scales of galaxies as a function of the redshift $z$.

The typical scales described above can be
also justified by carefully considering the quantum Jeans length which can be found by  inserting
the Madelung relation ~\cite{2011PhRvD..84d3531C,2014ASSP...38..107S},
$\psi(\bx,t)=\sqrt{\rho(\bx,t)}e^{iS(\bx,t)}$,
 into the SPE.
 Then, one can get an equation for density
 perturbation $\delta\rho\equiv \rho-\bar{\rho}$ \cite{2012A&A...537A.127C, Suarez:2011yf},
 \beq
 \label{pert}
  \frac{\partial^2 \delta\rho}{\partial t^2}+\frac{\hbar^2}{4m^2}\nabla^2 (\nabla^2 \delta \rho)
 - 4\pi G \bar{\rho}\delta\rho=0,
 \eeq
 which turns into
an equation for
  density contrast $\delta\equiv\delta \rho/\bar{\rho}=\sum_k \delta_k e^{ik\cdot \bx}$ with a wave vector $k$,
   \beq
  \frac{d^2 \delta_k}{d t^2} +  \left[c^2_q k^2-4\pi G \bar{\rho} \right]\delta_k=0,
 \eeq
  where $c_q=\hbar k/2m$ is a velocity
  for a mode with $k$.
By equating two terms in the parenthesis
~\cite{fuzzy}
   one can get
   the quantum Jeans length scale at a redshift $z$,
\beq
\label{lambdaQ}
\lambda_Q(z)= \left(\frac{\pi^3 \hbar^2 }{Gm^2\bar\rho(z)}\right)^{1/4},
\eeq
which is of the same order of $\bx_c$
in Eq. (\ref{dB2}).
The density perturbation with a size smaller than  $\lambda_Q$ cannot collapse
to form a galaxy.
Therefore, $\bx_c\simeq \lambda_Q$ represents the minimum
length scale  of galactic halos formed at $z$ ~\cite{Lee:2008ux,Lee:2015cos,Lee:2008jp},
 which can  resolve the small scale issues of the CDM
~\cite{corePeebles,PhysRevD.62.103517}.
Ref. \citealp{myhalo} showed that repulsive self-interaction of ULDM drastically increases
 the typical length scale, which is not considered in this work because the mechanisms are very different.
(See Ref. \citealp{chavanis} for more details.)
On the other hand,
it is well known that collision-less cold dark matter and self-interacting particle dark matter (not ULDM)
has a Jeans scale $v (\pi/G\bar\rho(z))^{1/2}$ which
is different from $\lambda_Q$ of ULDM. Here, $v$ is the velocity dispersion for cold dark matter and
 the sound velocity for self-interacting dark matter, respectively.
Therefore, ULDM has a unique time-dependency of its typical size different from those of other dark matter models.

It was shown~\cite{Lee:2015cos} that the observed size evolution of the most massive galaxies is consistent with our model,
if we use the relation $r_*(z)\propto \lambda_Q^{3/2}(z)\propto (1+z)^{-9/8}$ where $r_*$ is the size of the visible part of a galaxy.
Recently, a compact early galaxy referred to as RX J2129-z95  was observed ~\cite{2023Sci...380..416W} by James Webb space telescope.
It has stellar mass about $10^{7.63} M_\odot$  and its half-light radius is just about $16.2~pc$ at $z=9.51$,
which is much smaller than the typical size ($O(10^2)~ pc$) of similar mass galaxies at present.
This observation is also roughly consistent with our model.

From $\lambda_Q$ the quantum Jeans mass can be derived as
\beq
\label{MJ}
M_J(z)
=\frac{4}{3}
 \pi^{\frac{13}{4}}\left(\frac{\hbar}{G^{\frac{1}{2}}   m}\right)^{\frac{3}{2}} {\bar\rho}(z)^\frac{1}{4},
\eeq
which is of the same order of
$M$ in Eq. (\ref{M2}).
Because the structure formation theory of the ULDM model naturally supports the scales  $\bx_c$ and $M$, it also supports the derived scales  $v_c$, $a_c$ and $L_c$.

\begin{figure}[tbp]
\includegraphics[width=0.5\textwidth]{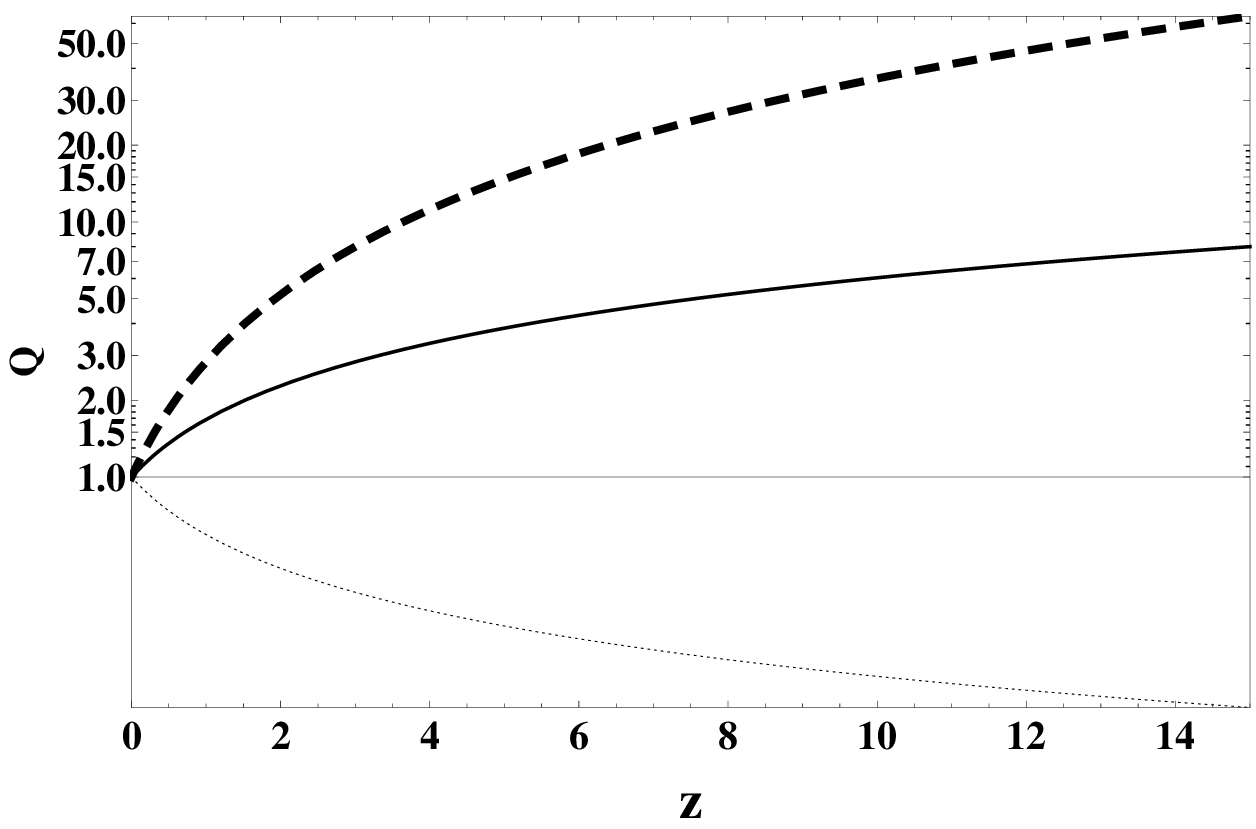}
\caption{ The evolution of the various physical quantities $Q$ of galaxies versus  redshift $z$, normalized to the current values. The dashed line corresponds to $Q=\psi_c$ and $Q=V_c$,  while the solid line represents $Q=M$ and $Q=L_c$, and the dotted line represents $Q=\bx_c$.}
\label{rarfig1}
\end{figure}

Not every scale is time dependent in the ULDM model. To see this
let us consider another popular  scaling  relation ~\cite{myhalo} which is useful for
relativistic cases;

\begin{align}    \label{}
t & \equiv   t_c
\, \hat{t}   = \frac{\hbar}{mc^2}  \hat{t}, \\
\bx &  \equiv  \bx_c %\alpha_s
\, \hat{\bx}   = \frac{\hbar}{mc}  \, \hat{\bx},\\
\psi & \equiv   \psi_c \, \hat{\psi}   = \frac{mc^2}{\hbar \sqrt{4\pi G}}   \hat{\psi}, \\
V& \equiv  V_c \, \hat{V}   = c^2 \hat{V},
\end{align}
where $\bx_c$ now corresponds to the Compton wavelength of the ULDM particles, and we use a different normalization
$4\pi\int |\psi|^2 d^3\bx=M$.

Equating $\bx_c$ with
the gravitational radius again one can get a typical mass scale
\beq
M=\frac{\hbar c}{mG}=1.3\times 10^{12} M_\odot \left( \frac{10^{-22}eV}{m}\right),
\eeq
which is about the maximum mass a boson star (ULDM soliton) can have,
 beyond which it will collapse to form a black hole ~\cite{myhalo, Khlopov1985}.
 Therefore, this theory also predicts
 the presence of super massive black holes with mass
 larger than $10^{12}M_\odot$ in the early universe. Interestingly, this bound is similar to the observed mass bound of galaxies.

The typical density for this normalization is
\beq
\rho_c=
\frac{c^4 m^2}{G {\hbar}^2}=
5.1\times 10^{15}~M_\odot/pc^3~\left(\frac{m}{10^{-22}eV}\right)^2.
\eeq

With this parameterization, obviously, $v_c=\bx_c/t_c=c,$ and $ L_c=M \bx_c v_c=N \hbar$.
Note that all these scales are independent of $\bar{\rho}$ and
time independent contrary to the scales in Eq. (\ref{scaling1}).

The probability current is now
\beq
J_c=\frac{c^5 m^2}{4\pi G\hbar^2}
= 1.24 \times 10^{14} M_\odot /pc^{2}/year \left(\frac{m}{10^{-22}eV}\right)^3,
\eeq
which is actually a mass flux, due to the new normalization $4\pi\int |\psi|^2 d^3\bx=M$.

The typical acceleration in this case
\beq
a_c= {c^3 m}/{\hbar}=45.5~ meter/s^2  \left( \frac{m}{10^{-22}eV}\right) ,
\eeq
which is the maximum acceleration scale
the soliton can have before
the soliton collapses to form a black hole and is different from the MOND scale. One can see that the gravitational acceleration of ULDM halos cannot exceed the limit without including other matter.
This upper bound of acceleration is one of the unique feature of ULDM, and any stable ULDM dominated structures should satisfy. If we use  Planck's constant $h$ instead of $\hbar$, $a_c$ becomes $7.24~meter/s^2$, which is similar to the gravitational acceleration of the earth.

\section{Discussions}

 Based on simple arguments using the SPE, we showed that one can easily derive the characteristic scale of various physical quantities for observed galaxies from quantum mechanics, if dark matter is ULDM.
The presence of the quantum scales is a very unique feature of the ULDM model.  For example,
The angular momentum per a DM particle is of order of $\hbar$, which means that the collective wave functions describing the rotating DM halos should be one of eigenstates of the angular momentum operators, because the ULDM particles are bosons in a single coherent state.
The maximum acceleration scale in
the ULDM model is another example
which is absent in other `classical' dark matter model such as CDM or warm dark matter.

Future data from observations such as the James Webb Space Telescope and the North American Nanohertz Observatory for Gravitational Waves (NANOGrav) ~\cite{NANOGrav:2023hfp} will open new windows into the galaxy evolution, possibly confirming these characteristic scales and their evolution.
Some of these scales have already been checked by numerical simulations, but more sophisticated simulations with baryons and black holes are needed to compare the theory to observations more precisely.

\subsection*{acknowledgments}
This work was supported by the Jungwon University Research
Grant(2021-031).
%
%\bibliographystyle{woc}
%\bibliography{sfdm}

\end{document}